# A GENERIC SURFACE MICROMACHINING MODULE FOR MEMS HERMETIC PACKAGING AT TEMPERATURES BELOW 200 °C

*R. Hellín Rico[1,2], J-P. Celis[2], K. Baert[1], C. Van Hoof[1] and A. Witvrouw[1]*

[1]Interuniversitair Micro-Elektronica Centrum, IMEC, B-3001, Leuven
[2]Katholieke Universiteit Leuven, Inst. MTM, B-3001, Leuven

## ABSTRACT

This paper presents the different processing steps of a new generic surface micromachining module for MEMS hermetic packaging at temperatures around 180°C based on nickel plating and photoresist sacrificial layers. The advantages of thin film caps are the reduced thickness and area consumption and the promise of being a low-cost batch process. Moreover, sealing happens by a reflow technique, giving the freedom of choosing the pressure and atmosphere inside the cavity. Sacrificial etch holes are situated above the device allowing shorter release times compared to the state-of-the-art.
With the so-called over-plating process, small etch holes can be created in the membrane without the need of expensive lithography tools. The etch holes in the membrane have been shown to be sufficiently small to block the sealing material to pass through, but still large enough to enable an efficient release.

## 1. INTRODUCTION

Unlike CMOS chips, chips containing MEMS (micro-electro-mechanical-system) cannot be directly packaged in a plastic or ceramic package (the so-called first level package) as MEMS are often composed of fragile and/or mobile free-standing parts that can easily be harmed during dicing and assembly. To avoid this damage, a MEMS device should be protected on the wafer level, before dicing.
This is possible with the so-called zero-level package [1]. Zero-level packaging is typically done by bonding a capping wafer or die to the MEMS wafer [2]. One route is to bond a micromachined capping wafer (usually glass or silicon) directly to the device substrate. However in that case, high process temperature techniques are required [3]. Other common ways are by using anodic or glass frit bonding. These are preferred because they require lower processing temperature (300-500 °C) than fusion bonding (1000°C). In addition, there has been a considerable fast development in packaging technologies using low temperature wafer bonding [4, 5, 6]. However, all the above mentioned techniques increase die area, and thus the cost substantially.

Thin film encapsulation is an alternative wafer-level packaging technique with a minimum amount of wasted area [7, 8]. By this technique hermetic encapsulation can be achieved by the fabrication and sealing of surface micromachined membranes covering the MEMS. By making the membrane thick enough or by using supports, plastic packaging can still be used as first-level packaging technique [9, 10].

In this work, a new generic surface micromachining module for MEMS thin film packaging with a metallic membrane at temperatures below 200 °C based on nickel electroplating is presented. With this technology the MEMS devices can be hermetically sealed and enclosed in a controlled atmosphere and pressure as required for proper operation and ensured lifetime of the MEMS device. The advantages of this integrated packaging technique are the reduced thickness and area consumption, the promise of being a low cost process and the low thermal budget.

State-of-the-art sealing is in general done using horizontal sacrificial etch channels. For example, Stark and Najafi reported surface micromachined caps with horizontal etch channels using nickel electroplating and solder sealing. Release times of several hours were needed [11]. In order to have a high-speed sacrificial release, sacrificial etch holes in the membrane are favourably situated above the device [12] (see Fig. 1). Recently, Rusu *et al.* reported a versatile sealing method for sealing vertical access holes by using a two layer thin film reflow process. Also in this work sealing is accomplished by using a reflow technique, but at much lower temperatures (180 °C) compared to Rusu *et al.* (600 °C) [7].

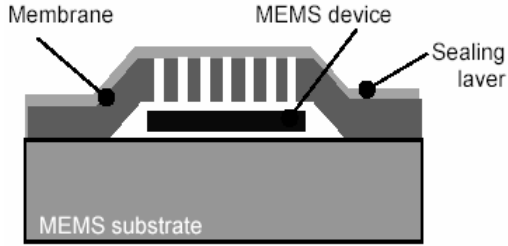

*Figure 1.* Schematic view of a MEMS package with access holes situated above the device

## 2. REQUIREMENTS

### 2.1 Membrane

The membrane layer needs to be rigid, strong and it preferably has a low tensile stress in order to prevent any bending or breaking. A high deposition speed is preferred to be able to deposit thick layers in case plastic moulding is required afterwards. We choose a nickel electroplated membrane on top of a sacrificial photoresist layer because of its high young's modulus (182GPa) [13]. The membrane layer will be structured in order to allow the sacrificial etching of the underlying material. Sacrificial layer access holes are situated above the device (see Fig. 1).

### 2.2 Sealing layer

The basic idea of the proposed sealing technique is to deposit a material with low melting temperature on top of openings in the membrane layer (see Fig. 2, step 7). After sacrificial layer etching, this material is then reflowed in a furnace with controlled atmosphere and pressure to close the final openings (see Fig. 2, steps 8 and 9). An optimization of the thickness of both membrane and sealing layer as well as an optimization of the size and shape of the etch holes is required. These holes should be large enough to enable efficient sacrificial etching but also small enough such that no or only negligible deposition inside the MEMS cavity takes place during sealing. The chosen sealing layer is indium with its low melting temperature (156.61 °C) because it allows a low thermal budget sealing process. Indium has also an excellent ductility which allows joining materials of different thermal expansion coefficients.

## 3. EXPERIMENTAL PROCEDURE

The process flow in this work starts with unreleased MEMS devices (Fig. 2, step 2), followed by the deposition and patterning of the photoresist sacrificial layer (sacrificial spacer) (Fig. 2, step 3). Subsequently, the seedlayer for plating is sputtered on (Fig. 2, step 4) and the non-conductive plating mould (photoresist) for the membrane is spun on. In order to reduce release times, etch holes are defined above the device by means of photolithography (Fig. 2, step 5). Then nickel is electroplated until a certain thickness, which is either smaller or larger than the mould resist height (Fig. 2, step 6). In the latter case, with the so-called over-plating process [14], smaller etch holes can be created in the membrane than in the former case. Then, a wetting layer and a low melting temperature sealing material are electrodeposited on the patterned membrane for further encapsulation of the MEMS device (Fig. 2, step 7). After mould and local seedlayer removal, the sacrificial release takes place by etching away the sacrificial layers of both MEMS device and thin film encapsulation layer (Fig. 2, step 8). Then the sealing material is reflowed to close the openings in the membrane (Fig. 2, step 9). Figure 2 shows the proposed packaging process schematically.

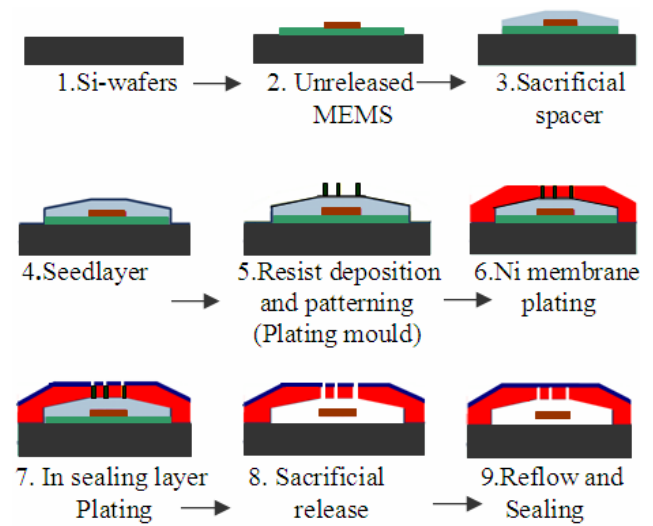

*Figure. 2* Process in steps for a generic module for MEMS hermetic packaging at T<200 °C

### 3.1 Membrane deposition and characterization

As mentioned previously, electroplated nickel was selected as the membrane material. The nickel is electroplated from a conventional sulfamate plating bath (from Shipley Ronal). According to Shipley's specifications, the pH in the solution was kept within the range of 3.8 to 4.5. The plating bath contains nickel sulfamate together with nickel chloride and boric acid. Boric acid buffers pH variations and nickel chloride is used to improve conductivity. The solution, which is held at a fixed temperature, is agitated during plating.

The stress in the membrane layer is a very important parameter for obtaining a stable membrane. Preferably the nickel membrane should have a low tensile stress at the end of the fabrication method. The two types of stress in thin films are thermal and intrinsic stress. Intrinsic stress is generated during film formation and is strongly dependent on process conditions. Thermal stress is caused by the difference in thermal-mechanical properties between the film and substrate and is unavoidable when the films are deposited at elevated temperatures. A common way to determine the stress ($\sigma$) in a film is by measuring the wafer curvature before and after film deposition and using Stoney's equation [eq. 1]. The substrates used in the experiments to optimize the stress in the nickel membrane were 6" silicon wafers with a seedlayer on top. This conducting layer was made by sputtering a Ti adhesion layer and a Cu seedlayer. After sputtering the seedlayer, the wafer bow is measured with a MX203 stress meter. The nickel film is then electroplated and the wafer bow is measured again. Then the stress is calculated using Stoney's equation:

$$\sigma = \frac{1}{6}\frac{Y_S}{(1-\upsilon_S)}\frac{t_S^2}{t_f}\left(\frac{1}{R_{after}} - \frac{1}{R_{before}}\right) \quad [\text{Eq. 1}]$$

where $Y_s$ is the Young's modulus of the substrate, $\upsilon_s$ is Poisson's ratio for the substrate, $t_s$ the substrate thickness and $t_f$ the film thickness. $R_{after}$ and $R_{before}$ are the radii of curvature after and before, respectively, plating the nickel.

### 3.2 Sealing layer deposition and characterization

As mentioned before, the sealing material should have a low melting temperature such that it can reflow at low temperatures. It is preferably deposited by a selective deposition technique in order to avoid material where this is not desired. Among the possible materials and techniques, electroplated indium was selected as sealing material. One more requirement for the sealing layer is that the perforation holes in the nickel membrane layer, that are needed for the sacrificial release, are sealed after reflow in such a way that the MEMS device underneath is not damaged by deposition of material through the holes. The sealing layer thickness must be large enough to seal these holes.

Using test wafers with patterned nickel layers, films of different thickness of indium were deposited by electroplating until the trenches were nearly closed. The indium anode had a purity exceeding 99.998 %. Normally after mould resist removal, seedlayer etch and sacrificial release, the reflow of the indium layer is done. On the test wafers, which are used to characterize and optimize the sealing process, no sacrificial layers were present. Therefore only the mould resist was removed before reflowing the indium. This reflow was performed in a Rapid Thermal Process (RTP) furnace in a $N_2$ atmosphere at atmospheric pressure and at temperatures above the melting temperature of the material.

### 3.3 Sacrificial layer removal

Etch test structures such as cantilevers were used to determine the etch rate of the photo-resist sacrificial layers in an oxygen plasma. The samples had a 6μm thick Ni structural layer and a 3μm thick photo-resist sacrificial layer. The samples were placed in a chamber with downstream oxygen plasma for a specific time at a pressure of 1600 mTorr and a chuck temperature of 25°C.

## 4. RESULTS AND DISCUSSION

### 4.1 Stress in nickel membrane

Nickel was plated with different thicknesses (6, 8 and 10 μm) at current densities (J) ranging from 1 to 4 A/dm$^2$ and at temperatures of 50ºC, 55ºC and 60ºC. Figure 3 shows the measured stress of nickel layers plated at a constant current density for different temperatures. From the results it was concluded that for the chosen sulfamate bath and titanium-copper seedlayer, the best results are observed for a plating current density at 1.41 A/dm$^2$ and a bath temperature of 55 °C. The cathode efficiency was always above 95%. The stress values obtained were always around 50 MPa compressive. Moreover, it was found that stresses are not dependent on the thickness grown. The effect of the reflow process on the internal stress of the plated nickel layers was studied by annealing nickel samples in a furnace with a controlled atmosphere at 180 °C for different times. These conditions reflect the real reflow process. The stress in the nickel layer was found to change from low compressive to low tensile after the reflow process (see Table 1).

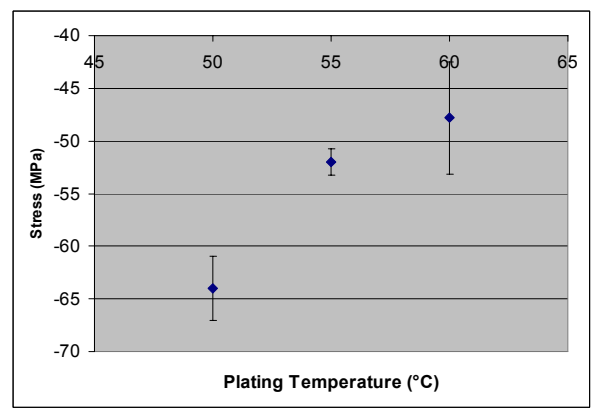

*Figure 3.* Stress of 6 μm thick Ni layers versus plating temperature. J=1.41 A/dm$^2$. Negative readings are compressive.

R. Hellin Rico, J-P. Celis, K. Baert, C. Van Hoof and A. Witvrouw
*A generic module for MEMS hermetic packaging at temperatures below 200 °C*

| Thickness plated (μm) | Average Stress (MPa) | Avg. Stress after annealing (MPa) |
|---|---|---|
| 6 μm | -50.3±1.5 | 91±2 |
| 8 μm | -52.8±0.3 | 81±1 |
| 10 μm | -45±2 | 85±3 |

*Table 1. Stress in plated Ni layers before and after annealing. The Ni was plated with $J=1.41$ A/dm$^2$ and $T=55°C$. Annealing was done at 180 °C for 5 min in RTP. Negative values indicate a compressive stress and positive values a tensile stress.*

The low tensile stress after annealing is exactly what is required for the membrane layer.

**4.2 Plating of membrane and sealing layer**

*4.2.1 Plating of the membrane*
Tests on nickel over-plating are done by filling high aspect ratio trenches. The process starts with 6" silicon wafers on which a titanium/copper seedlayer was deposited. To simulate the packaging process, trenches, which are defined by photolithography, mimic the access holes (see Fig. 2, step 5 and Fig. 3). The photoresist features consist of 6 μm thick and 1 to 15 μm wide lines and equally wide spaces (see Fig. 4). Then, nickel is electroplated into the exposed regions up to a thickness, which is larger than the resist thickness. Different thicknesses (8, 9 and 11 μm) were plated.

With this overplating process, small etch holes can be created in the membrane without the need of expensive lithography tools. For example, after electroplating an 11 μm thick nickel layer, features smaller than 10 μm are closed whereas larger features have their apertures reduced considerably (see Fig. 5).

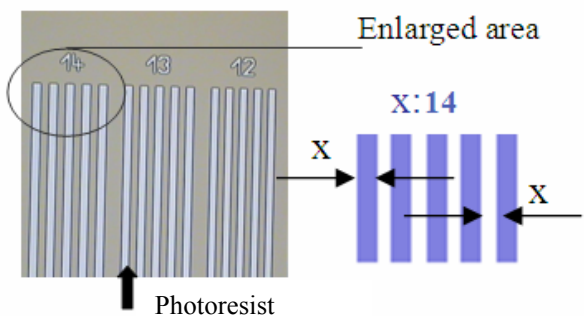

*Figure 4. Optical image of the resist litho-lines used as mould for plating. The enlarged feature consists of 14 μm wide lines with 14 μm gaps.*

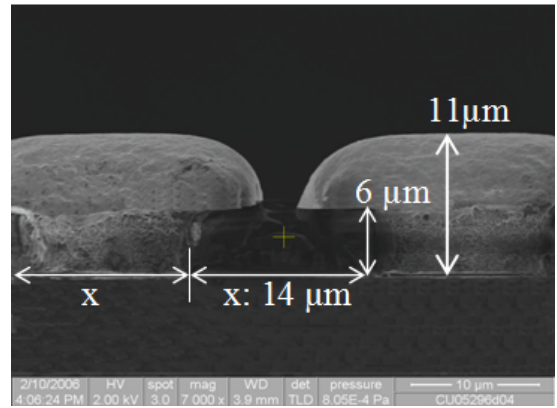

*Figure 5. 11μm high over-plated nickel (mushrooms). Features smaller than 10 μm wide by 6 μm thick are closed.*

*4.2.2 Plating of the sealing layer*
Prior to mould resist removal, the sealing layer is electrodeposited on the nickel trenches (see Fig. 2, step 4). Indium was plated with different thicknesses (1, 2, 3 and 4 μm) at current densities (J) ranging from 2 to 5 A/dm$^2$. Figure 6 shows a scheme of the indium plated above the nickel structures before and after reflow.

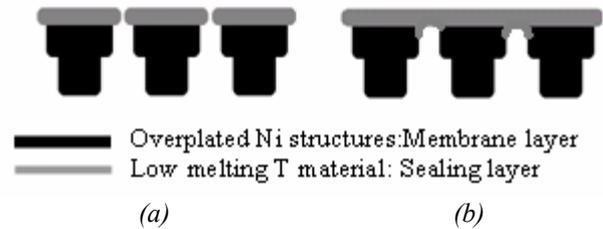

*Figure 6. One-layer sealing process:*
*(a) Deposition of a low-temperature melting material,*
*(b) Closing of the openings at an arbitrary chosen pressure and atmosphere*

**4.3 Sacrificial release process**

According to the real process, after plating the sealing layer, the mould resist is removed and the seedlayer is etched away. Then the sacrificial resist spacer below the Ni membrane needs to be removed. This can be done using an oxygen plasma.
The sacrificial etch rate was measured on test structures by evaluating the width of the released cantilevers after etching for a certain time. It was found to be 0.58μm/min (see Fig. 7 and 8). The underetching of the cantilever structures is almost linear as function of the etching time. From the measured etch rate, the release time for a certain design can be estimated. For example, if the distance in between etch holes is 8 μm, we expect a release time less than 10 minutes, while if the distance is 26 μm, the expected release time is less than half an hour.

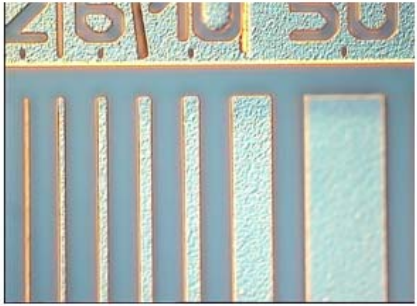

*Figure 7. Released cantilevers after 25 min in oxygen plasma: 2, 4, 6, 8, 10 and 25 µm width.*

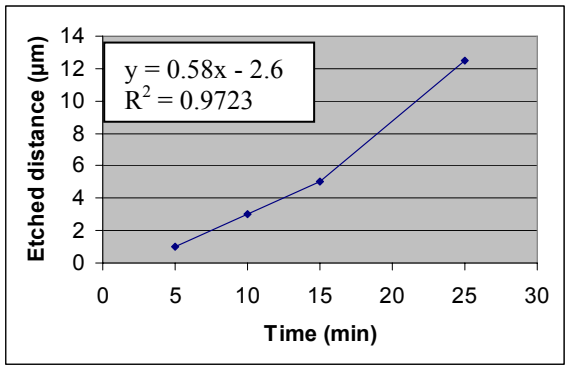

*Figure 8. Sacrificial etch rate measured by evaluating the width of the released cantilevers*

### 4.4 Reflow of the sealing layer

After etching away the sacrificial layers in a full process, the main goal of the sealing layer reflow is to close the membrane openings while keeping the cavity free of deposits. After optimizing the indium thickness and reflow conditions, openings as wide as 15µm and 6µm high on the test wafers with the nickel trenches were sealed. No deposited material was observed at the bottom of the trench (see Fig. 9 and 10).

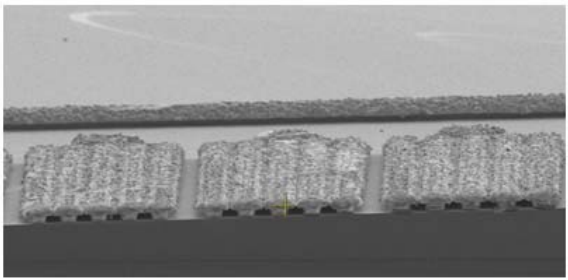

*Figure 9. Overview of the 15, 14 and 13 µm wide plated nickel trenches sealed after indium reflow.*

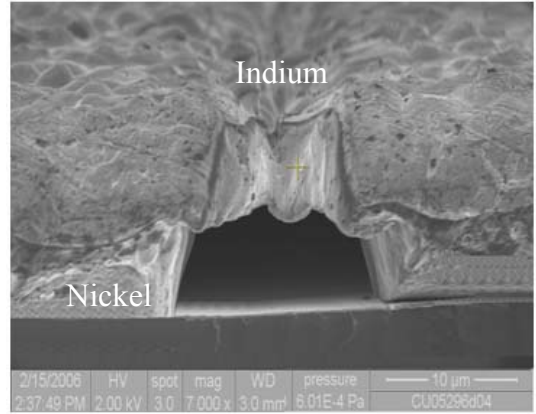

*Figure 10. SEM cross section image of 15µm wide by 11µm high Ni plated trenches sealed after indium reflow. No material was deposited in the cavities.*

### 5. CONCLUSION

In conclusion, we have developed all the different steps of a new generic surface micromachining based process for packaging MEMS devices at the wafer level at temperatures as low as 180°C (50°C lower that the state of the art for hermetic packages). The developed technique allows short release times because the holes are positioned above the device. After sealing the etch holes, no material was found to deposit in the cavities.
Small etch holes can be created in a thick membrane layer without the need of expensive lithography tools by controlling the membrane overplating process.

### ACKNOWLEDGEMENTS

The authors would like to thank Bert Du Bois for helpful discussions and for SEM analysis.